\documentclass[aps, pre, twocolumn, a4paper, longbibliography]{revtex4-1}
\usepackage{graphicx}
\usepackage{amsmath,mathtools}
\usepackage{placeins}
\usepackage{color}
\usepackage[utf8]{inputenc}

\usepackage{hyperref}
\usepackage{listings}
\usepackage{array}

\newcommand{\scriptrm}[1]{\mathrm{#1}}
\newcommand{\parm}[2]{#1_\scriptrm{#2}}
\newcommand{\twoparm}[3]{#1^\scriptrm{#3}_\scriptrm{#2}}
\newcommand{\qthr}[0]{\parm{q}{thr}}
\newcommand{\qm}[0]{q_m}

\newcommand{\qmaxfree}[0]{\twoparm{q}{max}{(free)}}
\newcommand{\rhomaxfree}[0]{\twoparm{\rho}{max}{(free)}}
\newcommand{\vjam}[0]{\parm{v}{jam}}
\newcommand{\vminfree}[0]{\twoparm{v}{min}{(free)}}
\newcommand{\Pjam}[0]{\parm{P}{jam}}
\newcommand{\un}[1]{\,{\rm #1}}
\newcommand{\revision}[1]{#1}
\begin{document}

\title{The Importance of Antipersistence for Traffic Jams}
\author{Sebastian M. Krause}
\author{Lars Habel}
\author{Thomas Guhr}
\author{Michael Schreckenberg}
\affiliation{Faculty of Physics, University of Duisburg-Essen, 47058 Duisburg, Germany}

% \pacs{89.75.-k}{Complex systems}
% \pacs{05.40.Jc}{Brownian motion}
% \pacs{89.40.-a}{Transportation}

\begin{abstract}
Universal characteristics of road networks and traffic patterns can help to forecast and control traffic congestion. The antipersistence of traffic flow time series has been found for many data sets, but its relevance for congestion has been overseen. 
Based on empirical data from motorways in Germany, we study how antipersistence of traffic flow time-series impacts the duration of traffic congestion on a wide range of time scales. We find a large number of short lasting traffic jams, which implies a large risk for rear-end collisions.
\end{abstract}

\maketitle

\section{Introduction}
Intraurban road networks in agglomerations and megacities often operate near or above their designed specifications in terms of e.g. maximum capacities, which leads to congestion and increases travel times~\cite{oecd2007managing}. Exceeding the specifications can also result in an increased wear of important parts of the network infrastructure, in particular bridges. During subsequent maintenance works, the road capacities are typically reduced, which adds to the problem. Under these circumstances, road authorities are faced with the challenge of optimal traffic assignment and control. To this end, universal characteristics of road networks and the according traffic patterns~\cite{popovic2012geometric} can help to identify systemic bottlenecks~\cite{li2015percolation}. While local traffic time series are best characterised with identifying different traffic states and state transitions~\cite{persaud1998exploration,kerner2002empirical}, network aspects are well represented with fractal scaling laws~\cite{popovic2012geometric,petri2013entangled,barthelemy2011spatial}. Both aspects are grounded on empiric evidence in very diverse situations and are well understood with microscopic models. Especially the spatio-temporal behaviour of traffic patterns is explained comprehensively with the three-phase traffic theory~\cite{kerner2004physics,kerner2009introduction}, which distinguishes between free flow, synchronised traffic and wide moving jams. The latter two phases are summarised under the heading of congested traffic.

Empirical studies find fractal properties in local traffic time series~\cite{peng2010long,karlaftis2009memory,toledo2004modeling,shang2007fractal,li2007multifractal,wang2014traffic,kantelhardt2013phases,zaksek2015tgf}, also based on methods like detrended fluctuation analysis (DFA)~\cite{peng1994mosaic,kantelhardt2002multifractal}. These findings are consistent with results from cellular automata traffic flow models~\cite{wu2008long,zaksek2015tgf}. Fractal modelling based on fractional Brownian motion (fBm)~\cite{mandelbrot1968fractional} was used for forecasting traffic flow~\cite{vojak1994multifractal}. \revision{fBm is a generalisation of the Wiener process (also known as random walk). Its fractal character is a self-similarity of the time series. If time is stretched with factor $A$, the data is stretched with factor $A^H$, where the parameter $0<H<1$ is known as the Hurst exponent. For $H=1/2$, fBm simplifies to the diffusion-like behaviour of the Wiener process. For $H>1/2$, fBm is super-diffusive. Increments of the time series are long-term correlated which is called persistent behaviour. In this letter we are interested in the case $H<1/2$. Then we have sub-diffusive behaviour and anti-correlated increments, which is called anti-persistence. This implies large fluctuations on short time-scales which reverse fast. The implications of fractal time series} for traffic breakdown are not well understood up to now. This limits the implicative relevance of fractal properties for the fine-tuning of traffic models.

\begin{figure}
      \centering
      \includegraphics[width=0.9\columnwidth]{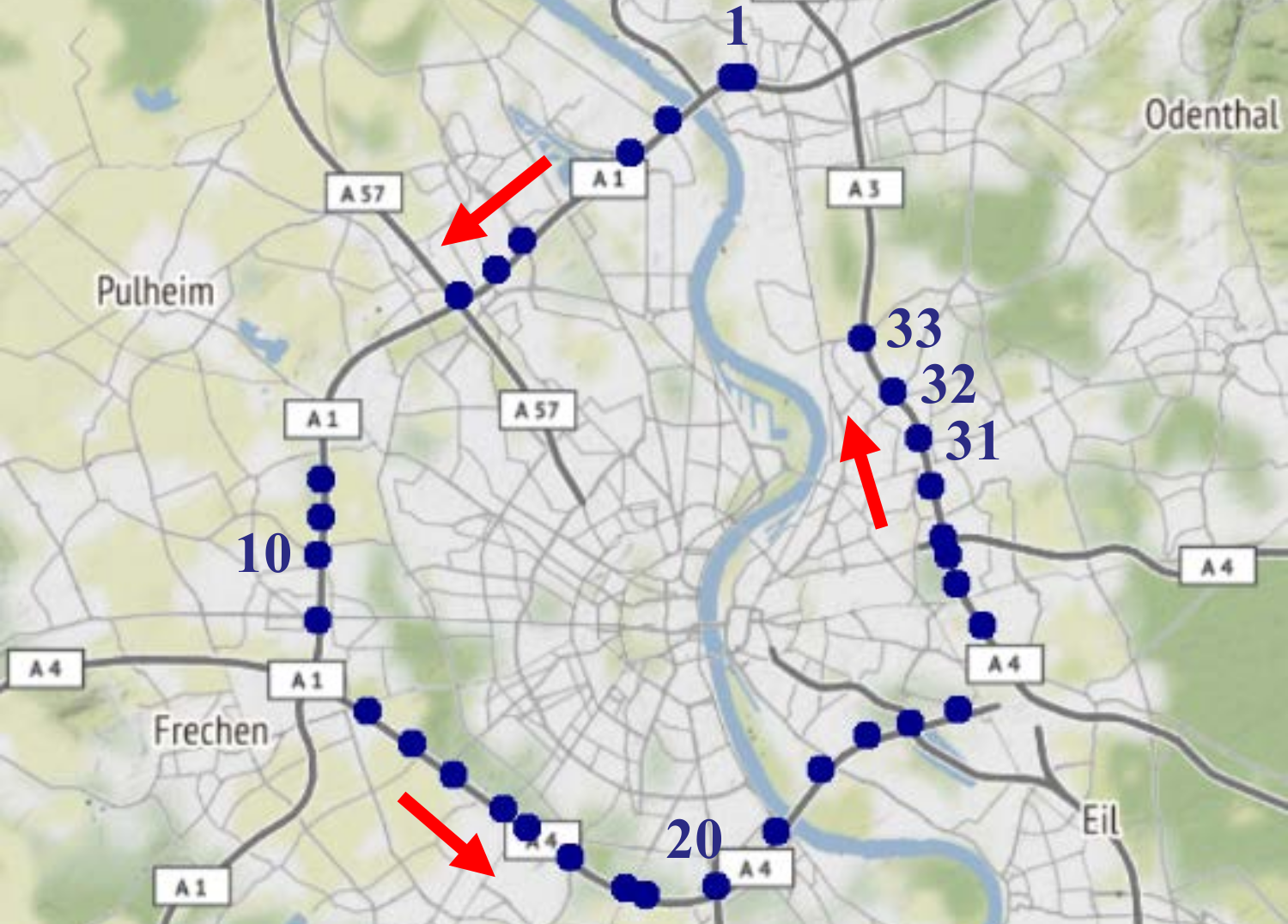}
      \caption{Locations of traffic detector crosssections $i$ (dots) on the Cologne orbital motorway. Specific crosssections are numbered counter-clockwise. Map tiles by Stamen Design, under CC BY 3.0. Data \copyright~OpenStreetMap Contributors.}
      \label{fig:coords}
\end{figure}

Here we study how the fractal nature of traffic flow time-series impacts the duration of traffic congestion. We show that the corresponding distribution is very broad. We succeed in explaining it as a consequence of antipersistence in traffic flow. For our empirical analysis, we use traffic data from inductive loops located at crosssections $i \in \{1,...,33\}$ on the Cologne orbital motorways A1, A3 and A4 in Germany, which are depicted in fig.~\ref{fig:coords}. Motorway traffic in this area has also been studied in~\cite{neubert1999single,lubashevsky2001long,belomestny2003complete}. The data set comprises traffic flows $q(i,t)$, densities $\rho(i,t)$ and velocities $v(i,t)$ \revision{averaged over 1-minute intervals} $t$ of the year 2015 and all crosssections $i$. A traffic flow $q(i,t)$ is defined as the number of vehicles passing the crosssection $i$ on all lanes in minute $t$, whereas $v(i,t)$ denotes the corresponding averaged vehicular velocities.

%Empirical origin of spontaneous and induced traffic breakdowns~\cite{kerner2015nuclei}

%Upstream propagation velocities of empirical traffic jams~\cite{rehborn2011empirical}

%non-recurrent congestion duration (accidents...) on Dutch highways~\cite{adler2013road}

%delays caused by work zones on Korean highways~\cite{chung2011assessment}

%\todo{
%synchronized flow stability~\cite{kerner1997experimental}

%distinct classes of time series~\cite{chrobok2004diff}}

\section{Statistics of traffic jams}

\begin{figure}[htb]
\begin{center}
    \includegraphics[trim=0 0 0 0,clip,width=1.0\columnwidth]{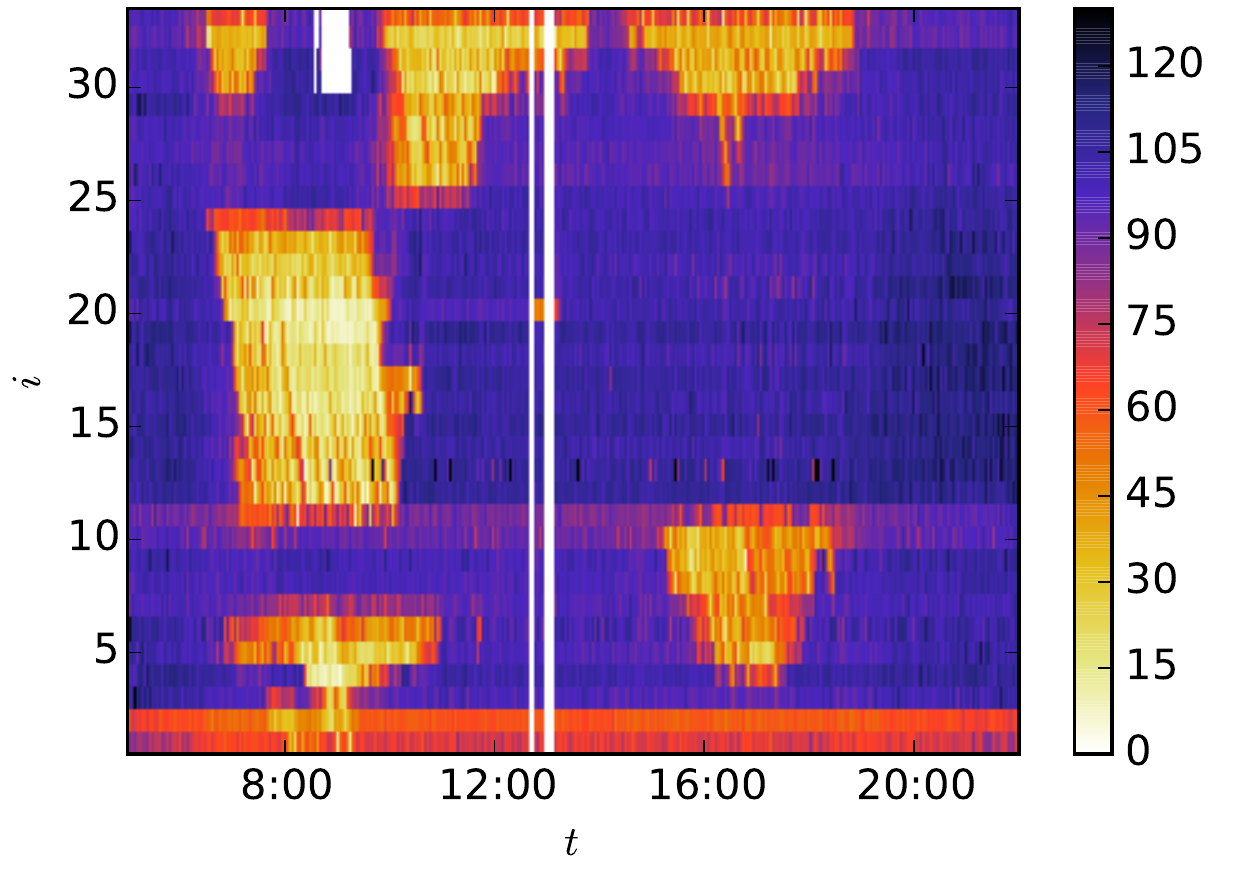}
    \caption{Velocity profile during Wednesday, 21 October 2015 along the crosssections $i$ specified in fig.~\ref{fig:coords}. The colours indicate vehicular velocities $v(i,t)$ in km/h, averaged over one minute windows on all possible lanes of a crosssection. The white regions due to missing data are skipped in the further data processing.}
    \label{fig:velocity}
\end{center}
\end{figure}

Fig.~\ref{fig:velocity} displays velocity profiles versus time for a specific day. At crosssections 1 and 2, there is a fixed speed limit of $60\un{km/h}$. Sections 3--7 and 12--25 have fixed speed limits of at most $120\un{km/h}$, whereas sections 8--11 and 26--33 are equipped with variable speed limit signs. For crosssections with numbers larger than 25, traffic jams develop before 8:00, around 12:00 and around 16:00. 

For understanding the spatio-temporal patterns shown in fig.~\ref{fig:velocity}, we use definitions from three-phase traffic theory. Free flow and congested traffic states can be distinguished by calculating a minimum velocity of free flow $\vminfree=\qmaxfree/\rhomaxfree$, where $\qmaxfree$ is the maximum free flow and $\rhomaxfree$ is the maximum free density~\cite{kerner2004physics}. Then, states with $v(i,t)<\vminfree$ are considered congested. However, this separation becomes erroneous where speed limits change often. To identify congested traffic, we therefore consider times and crosssections with $v(i,t)<\vjam$ below a fixed threshold velocity \revision{$\vjam=50\un{km/h}$} as congested. 

\begin{figure}[htb]
\begin{center}
    \includegraphics[width=1.0\columnwidth]{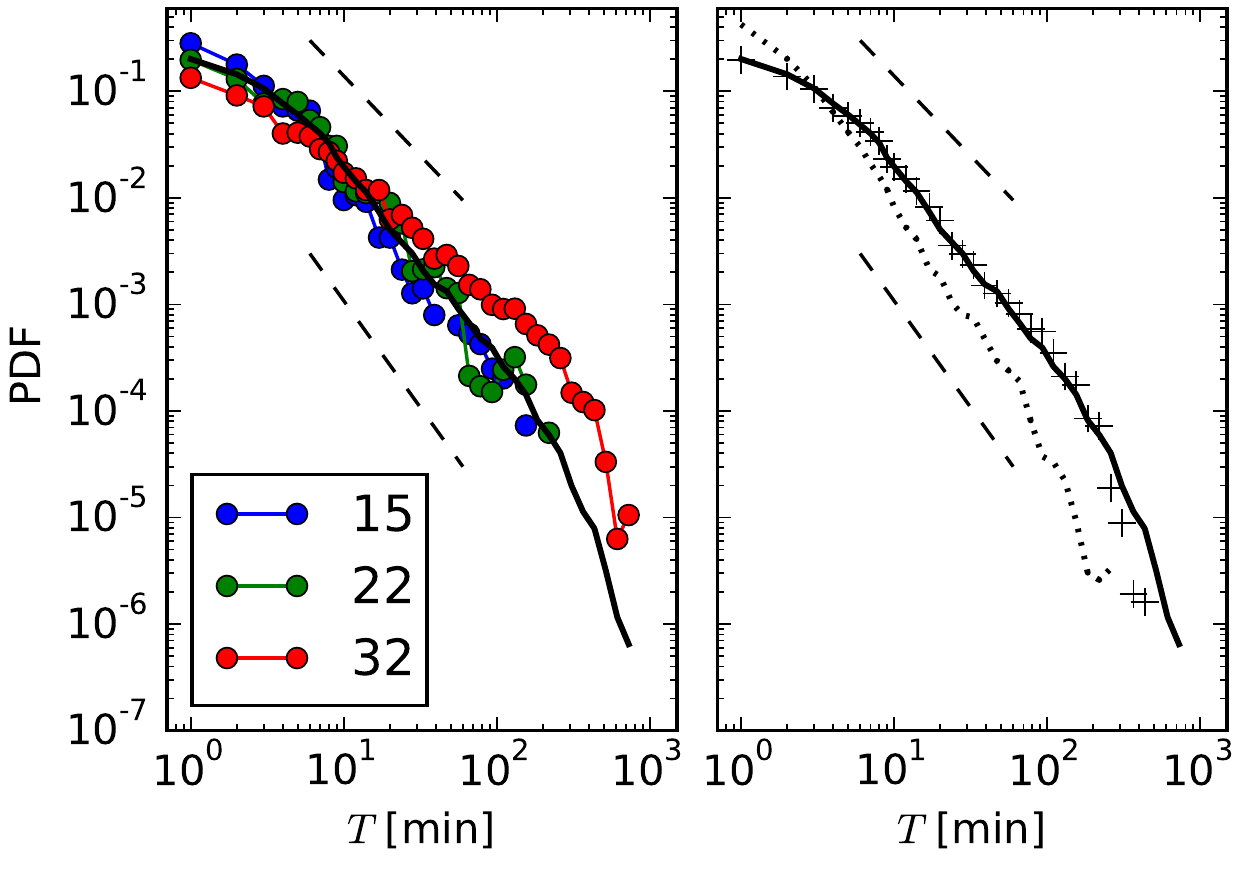}
    \caption{Left: PDF of traffic congestion durations $T$ for \revision{$\vjam=50\un{km/h}$} in double logarithmic plot. Symbols indicate different crosssections, the black solid line is the average over crosssections 3 to 33. For comparison, power laws $T^{-\gamma}$ with exponent $\gamma=3/2$ (upper dashed line) and $\gamma=2$ (lower dashed line) are shown. Right: The average result (black solid line) changes only slightly for data reduced to the first three months (crosses), or for reduced $\vjam=20\un{km/h}$ (dotted line).}
    \label{fig:length_}
\end{center}
\end{figure}

In fig.~\ref{fig:length_} we show the probability density function (PDF) of traffic congestion durations $T$. We identify a local congestion of duration $T$, if at a certain crosssection $i$ we have
\begin{align}
 &v(i,t)\geq \vjam\\
\textrm{and} \quad &v(i,t+T+1)\geq \vjam\\
\textrm{and} \quad &v(i,\tau)< \vjam
\end{align} for $t<\tau\leq t+T$. The resulting distribution is very broad. The dashed lines provide power laws $T^{-\gamma}$ for comparison, with $\gamma=3/2$ and $\gamma=2$. As shown, the results are qualitatively the same for different $\vjam$ as well as for data reduced to the first three months. Summarising, we find a robust power-law behaviour with exponents in the range $\gamma=3/2$ up to $\gamma=2$. \revision{The power law behavior starts at about $T=5\un{min}$.} A cutoff around 200 minutes results from the limited duration of rush hours. Importantly, the small exponent $\gamma$ implies that traffic congestion durations on all scales from minutes to hours are relevant. \revision{Overall, jams of duration $T<5\un{min}$ contribute about 8\% to the total sum of jam hours, jams with $5\un{min}\leq T \leq 10\un{min}$ add 11\%, jams with $10\un{min} < T < 100\un{min}$ add 44\%, and jams with $100\un{min}\leq T \leq 200\un{min}$ add 19\%. We concentrate on the power law regime of jam durations $5\un{min}\leq T \leq 200\un{min}$, as it spans almost two orders of magnitude and it describes how long lasting and short lasting jams relate to each other. For smaller exponent $\gamma$, the short lasting jams would be suppressed, while} for larger exponent $\gamma$, the long durations would be of minor importance. 

%For understanding congestion durations as a consequence of overcritical traffic flow, let us analyse flow time series and the reaction of the velocity to large flows.
To link congestion durations with traffic conditions leading to a traffic breakdown, we analyse traffic flow time-series and the reaction of the velocity on large flows. Traffic breakdowns occur at bottlenecks with some probability, if large traffic flows are present~\cite{kerner2004physics,kerner2009introduction}. 
%{\color{red}Begruending? Motivation? Road sections 32 and 33 are often the starting point of traffic jams. Traffic breakdowns occur at bottlenecks with some probability, if large traffic flows $q\geq\qcrit$ are present.} 
%It is interesting to investigate the time evolution of $q(i,t)$ prior to a traffic breakdown. As $\qcrit$ is difficult to find in a limited empirical data set, 
We use a fixed threshold flow $\qthr$ to calculate the breakdown probability $\Pjam(i,\qthr)$ with the following algorithm: Consider all events with $q(i,t)>\qthr$ and $v(i,t)\geq \vjam$, where in each of the following minutes $\Delta t \in \{1,\dots,5\}\un{min}$ it holds separately either
\begin{align}
 &v(i,t+\Delta t)<\vjam \textrm{ (jam occurs) }\\
\textrm{or} \quad  &v(i,t+\Delta t)\geq \vjam \textrm{ and } q(t+\Delta t)>\qthr.
\end{align}
Among these events, events with traffic breakdown have $v(i,t+\Delta t)<\vjam$ for at least one $\Delta t$, \revision{with $\Delta t$ after the traffic breakdown being ignored. The fraction of traffic breakdown events yields the breakdown probability $\Pjam$, where the minimum free flow $q$ in the considered time interval is restricted to $q_{\rm thr}\pm 2\un{min^{-1}}$.}
%%%
\begin{figure}[htb]
\begin{center}
    \includegraphics[width=1.0\columnwidth]{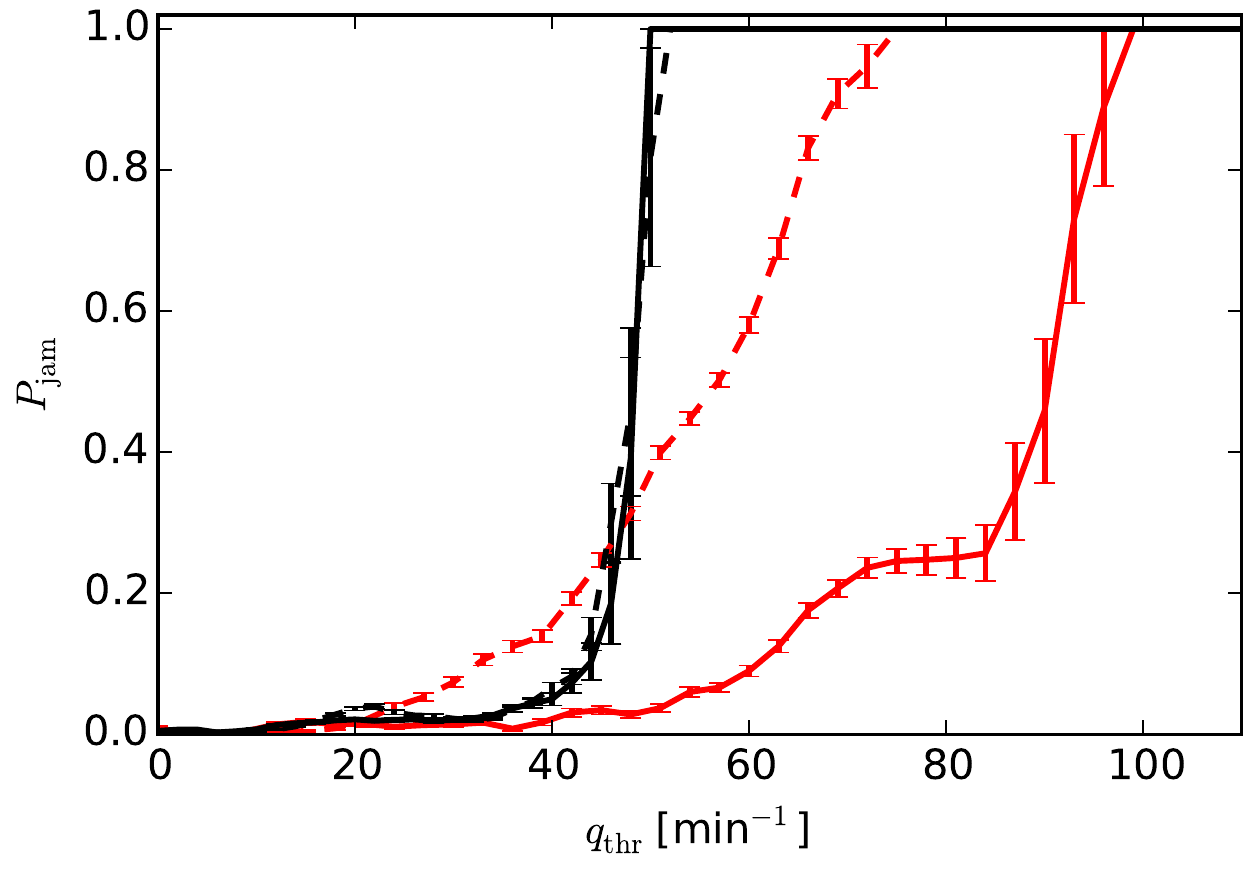}
    \caption{Probability $\Pjam$ that during up to five minutes with $q>\qthr$ the velocity falls below $\vjam$, versus threshold flow $\qthr$. Black lines are for crosssection 11 and red lines for crosssection 33. Line styles are \revision{solid for the days until 24 Mai 2015 and dashed for the remainder of the year. Error bars are based on the standard deviation of event counts, assuming Poisson statistics.} }
    \label{fig:p_jam}
\end{center}
\end{figure}
%%%

In fig.~\ref{fig:p_jam} we present resulting breakdown probabilities for crosssections 11 and 33 and varied threshold flow $\qthr$, split into time intervals \revision{until 23 May 2015 and starting from 27 May 2015.} For all curves, a sharp jump can be observed. The minimum flow with breakdown probability $\Pjam=1$ is denoted as $\qmaxfree$~\cite{kerner2009introduction}. \revision{We find $\qmaxfree$ in the range $50\un{min^{-1}}$ to $51\un{min^{-1}}$ for crosssection 11, and values from $74\pm 1\un{min^{-1}}$ up to $97\pm 2\un{min^{-1}}$ for crosssection 33.} The maximum free flow $\qmaxfree$ varies strongly between the crosssections, mainly because of the different number of lanes at each section. At crosssection 33, the maximum free flow reduces strongly \revision{in the second time interval because of a changed lane configuration at an on-moving construction site.} At other crosssections, the maximum free flow stays almost constant over the year.

\begin{figure}[htb]
\begin{center}
    \includegraphics[width=1.0\columnwidth]{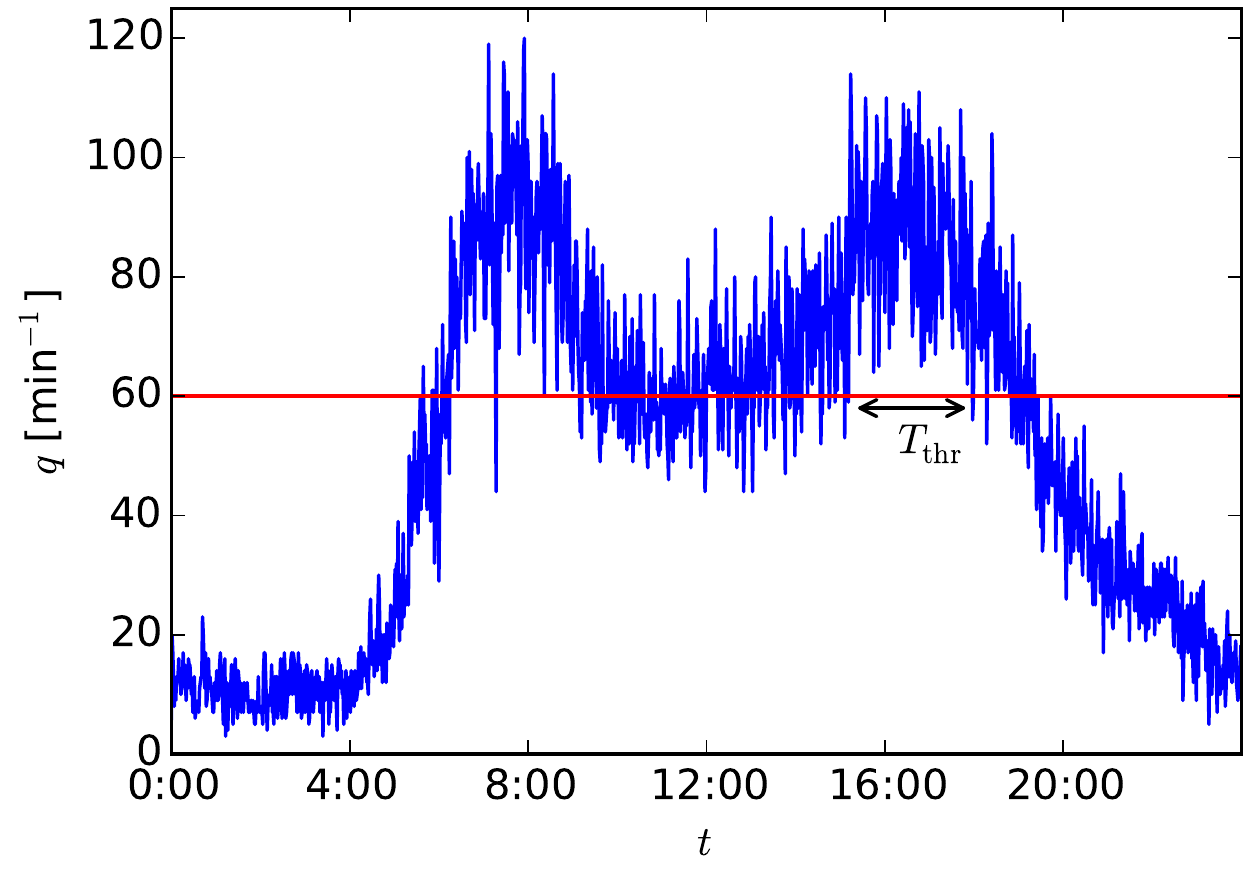}
    \caption{Traffic flow time series $q(t)$ for road section 22 on Tuesday, 14 July 2015. The duration $T_{\rm thr}$ of a period with $q>\qthr=60\un{min}^{-1}$ is indicated with a double arrow.}
    \label{fig:j_timeseries}
\end{center}
\end{figure}

\begin{figure}[htb]
\begin{center}
    \includegraphics[width=1.0\columnwidth]{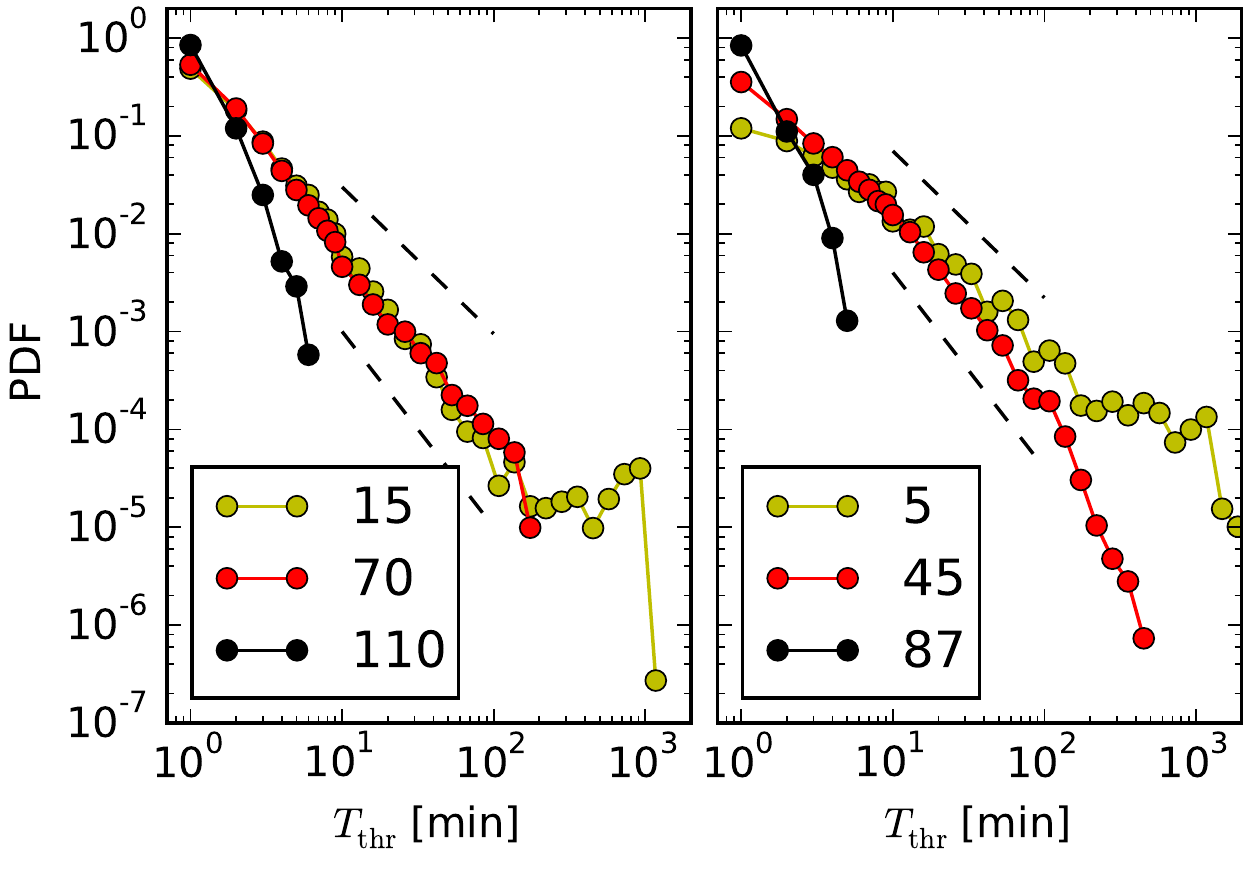}
    \caption{PDF of durations $T_{\rm thr}$, during which the flow is above a threshold, $q>\qthr$. The legends indicate different choices of $\qthr$. On the left, only days without detected congestion at crosssection 22 are considered. On the right, only days with at least four hours of congestion at crosssection 32 are considered. The dashed lines have exponents $\gamma_{\rm thr}=3/2$ and $\gamma_{\rm thr}=2$.}
    \label{fig:j_duration}
\end{center}
\end{figure}

Knowing that above a certain $\qthr$ traffic breakdown is likely to occur within a few minutes, we further analyse, for how long traffic flow exceeds $\qthr$, but does not break down~\cite{knorr2013high}. Would $\qmaxfree$ be reduced to the smaller flow $\qthr$ (for example due to construction works), traffic jams would occur as long as $q>\qthr$. 
In fig.~\ref{fig:j_timeseries}, traffic flow time series for Tuesday, 14 July 2015 are displayed. The time series shows strong fluctuations for short times, and a trend with one rush hour around 8:00 and a second rush hour around 16:00. Let us assume a threshold flow of $\qthr=60\un{min^{-1}}$, corresponding to the red line. We identify durations $T_{\rm thr}$ during which the flow exceeds a certain threshold $q_{\rm thr}$, i.e. $q>\qthr$. Due to the fluctuations in $q(t)$, we expect shortest durations $T_{\rm thr}$ down to a minute. In fig.~\ref{fig:j_timeseries} the largest duration $T_{\rm thr}$ is highlighted with the double arrow and spans almost three hours. The PDF of $T_{\rm thr}$ for different crosssections and thresholds $\qthr$ are shown in fig.~\ref{fig:j_duration}. On the left, we use \revision{250} days without a single minute of traffic jam in crosssection 22. \revision{For the threshold at the large flow $\qthr=110\un{min^{-1}}$ (black symbols), longer durations are not seen.} This is because here we restricted the data to days without traffic jam. Longer durations with such high flow would result in a traffic jam. For a smaller threshold $\qthr=70\un{min^{-1}}$ (red symbols) we see a power law distribution of durations $\propto \left(T_{\rm thr}\right)^{-\gamma_{\rm thr}}$, with exponent close to $\gamma_{\rm thr}=2$. Were the critical flow reduced to the smaller flow (for example due to reconstruction works), the durations $T_{\rm thr}$ would translate into traffic jam durations $T$. The PDF of traffic jam durations shows a power law with exponent close to $\gamma=2$, 
cf. fig.~\ref{fig:length_}. This result supports the interpretation that durations $T_{\rm thr}$ of the traffic flow $q(t)$ being above a threshold explain the distribution of traffic jam durations $T$. Only short traffic jam durations $T$ are suppressed compared to short $T_{\rm thr}$\revision{, meaning that the distribution of jam durations $T$ is reduced compared to power law behaviour for $T<5\un{min}$, see fig.~\ref{fig:length_}.} This is because the traffic needs some time to break down. \revision{Nevertheless, we already mentioned the strong contribution of short lasting jams even outside the power law regime.} For small threshold $\qthr=15\un{min^{-1}}$, the longest duration $T_{\rm thr}$ can be as long as the whole working day, resulting in a peak at about $600\un{min}$ or ten hours. To illustrate that these statistical features of traffic flow are not altered if the velocity breaks down, we consider \revision{149} days with at least four hours with $v<\vjam$ a day, in the second half of 2015 in crosssection 32. Results are shown on the right of fig.~\ref{fig:j_duration}.

\begin{figure}[t]
\begin{center}
    \includegraphics[width=\columnwidth]{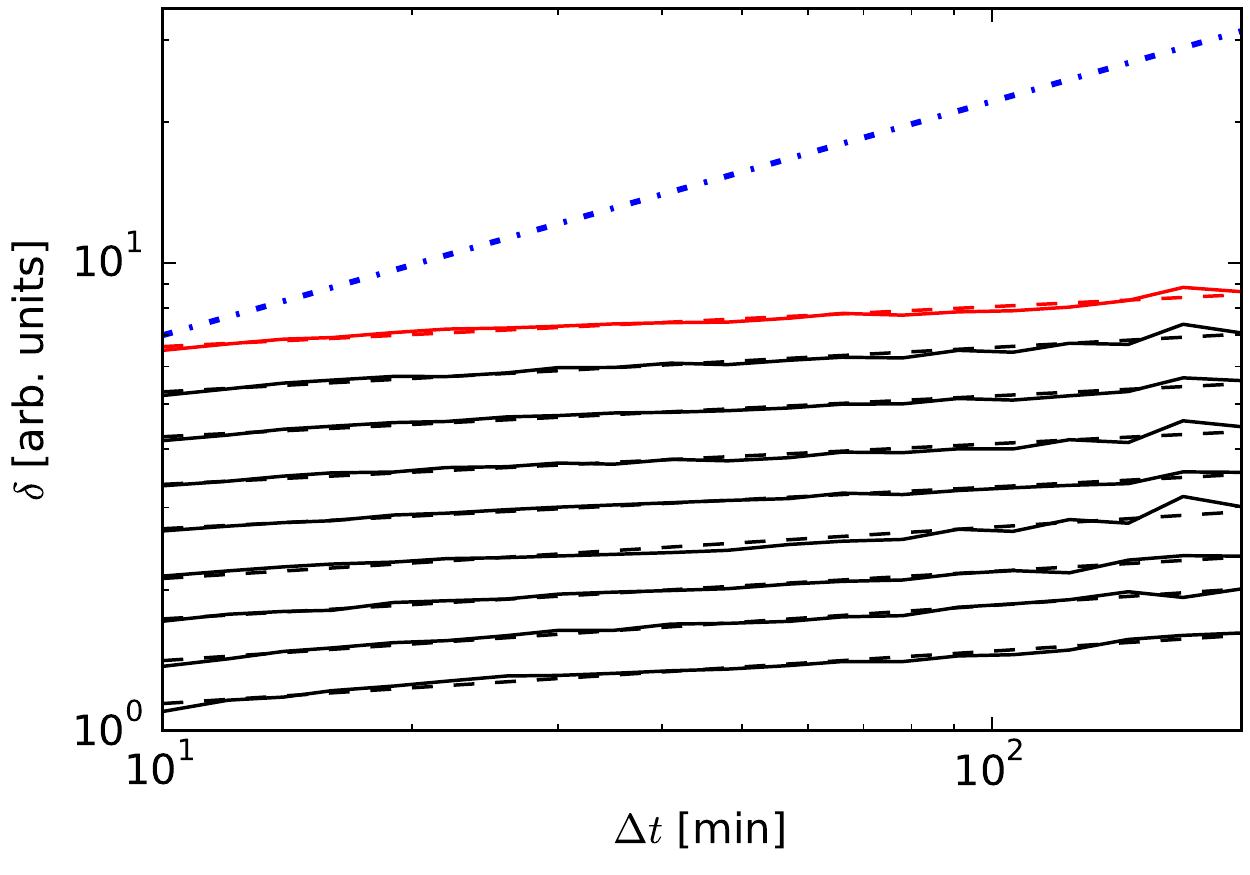}
    \caption{Detrended standard deviation $\delta$ of the flow time series $q(t)$ vs.~size of the sub-samples $\Delta t$. The red line corresponds to fig.~\ref{fig:j_timeseries}, black lines to flow time series on different times and crosssections. The blue dash-dotted line with exponent $1/2$ corresponds to Brownian motion.% Right: For every road section, the histogram of Hurst exponents is shown with a color map. Notice that we have varying Hurst exponents, one for every day.
    }
    \label{fig:dfa}
\end{center}
\end{figure}

\section{Explanation}
To understand the durations $T_{\rm thr}$ with $q>\qthr$, we compare them with fractional Brownian motion (fBm)~\cite{mandelbrot1968fractional}. We denote the fBm random function as $B_H(\tilde{t})$ with Hurst exponent $0<H<1$ and dimensionless time $\tilde{t}$. The defining property of the fBm with $B_H(0)=0$ is its dependency structure~\cite{mandelbrot1968fractional} for times $\tilde{t}$, $\tilde{s}\geq 0$,
\begin{equation}
    2\left<B_H(\tilde{t}) B_H(\tilde{s})\right> = \tilde{t}^{2H}+\tilde{s}^{2H}-|\tilde{t}-\tilde{s}|^{2H},\label{eq:dependence}
\end{equation}
where $\left<\phantom{B}\right>$ is the ensemble average over realisations. 

The PDF of durations $T_{\rm fBm}$ during which the time series exceeds a certain threshold $B_{\rm thr}$, i.e. $B_H(\tilde{t})>B_{\rm thr}$, is known to scale with a power law as $(T_{\rm fBm})^{-\gamma_{\rm fBm}}$ with $\gamma_{\rm fBm}=2-H$~\cite{ding1995distribution}. The traffic flow time series in fig.~\ref{fig:j_timeseries} shows strong fluctuations on short time scales, and thus anti\-per\-sistent, i.e. non-Markovian, behaviour with anti\-cor\-re\-lated increments and Hurst exponent $H<1/2$~\cite{mandelbrot1968fractional}. Another implication of anti\-persistent fBm is a subdiffusive behaviour, with variance increasing sub-linear in time as 
\begin{equation}
    \left<(B_H(\tilde{t}+\Delta \tilde{t})-B_H(\tilde{t}))^2\right> = |\Delta \tilde{t}|^{2H}.
\end{equation}
This result can be derived from eq.~\eqref{eq:dependence}. For small $H$ it implies that changes are large on short times and stagnating for longer times. The time dependence of the variance can be used for estimating $H$ from flow time series $q(t)$. To deal with the trend in the signal with pronounced rush hours, we use detrended fluctuation analysis (DFA)~\cite{peng1994mosaic,kantelhardt2002multifractal}. We divide the time series $q(t)$ of the day into sub-samples of length $\Delta t$ and correct the linear trend in each sample. Then we calculate the standard deviation in each sub-sample, and average over all sub-samples, to obtain the average standard deviation $\delta$. We repeat this procedure for different $\Delta t$. 
In fig.~\ref{fig:dfa} we show how the detrended standard deviation $\delta$ depends on the sub-sample size $\Delta t$. The red solid line corresponds to fig.~\ref{fig:j_timeseries}.
According to~\cite{kantelhardt2002multifractal}, the sub-sample size should be chosen larger than 10 elements and smaller than about $1/4$ of the full sample size. The Hurst exponent $H$ can be identified as the slope of the linear fit in the log-log plot~\cite{kantelhardt2002multifractal}. For the red curve we find $H=0.085$. Other examples for different days and crosssections 15 and 32 are shifted for better visibility. The dash-dotted line with exponent $1/2$ corresponds to Brownian motion. With $H<1/2$ we find strong subdiffusive behaviour in a range from ten minutes up to three hours. Performing DFA for single days on all crosssections, we find Hurst exponents between $0.038$ and $0.24$, with mean $0.088$ and standard deviation $0.028$. Days with more than ten minutes of missing data are neglected. For the Kerner-Klenov-Wolf cellular automaton three-phase traffic flow model, anti\-persistent behaviour of the free traffic density is reported in~\cite{wu2008long}. With the use of DFA, Hurst exponents down to $H=0.1$ are found in synthetic data. An analysis of real world data finds Hurst exponents around $H=0.17$ for free flow traffic~\cite{peng2010long}. Another study finds persistence in real traffic data, however the data is not detrended there~\cite{karlaftis2009memory}. %{\color{red} Our findings are consistent with correlated distances of single vehicle data in free traffic~\cite{neubert1999single,knospe2002single}.} 

\begin{figure}[t]
\begin{center}
    \includegraphics[width=\columnwidth]{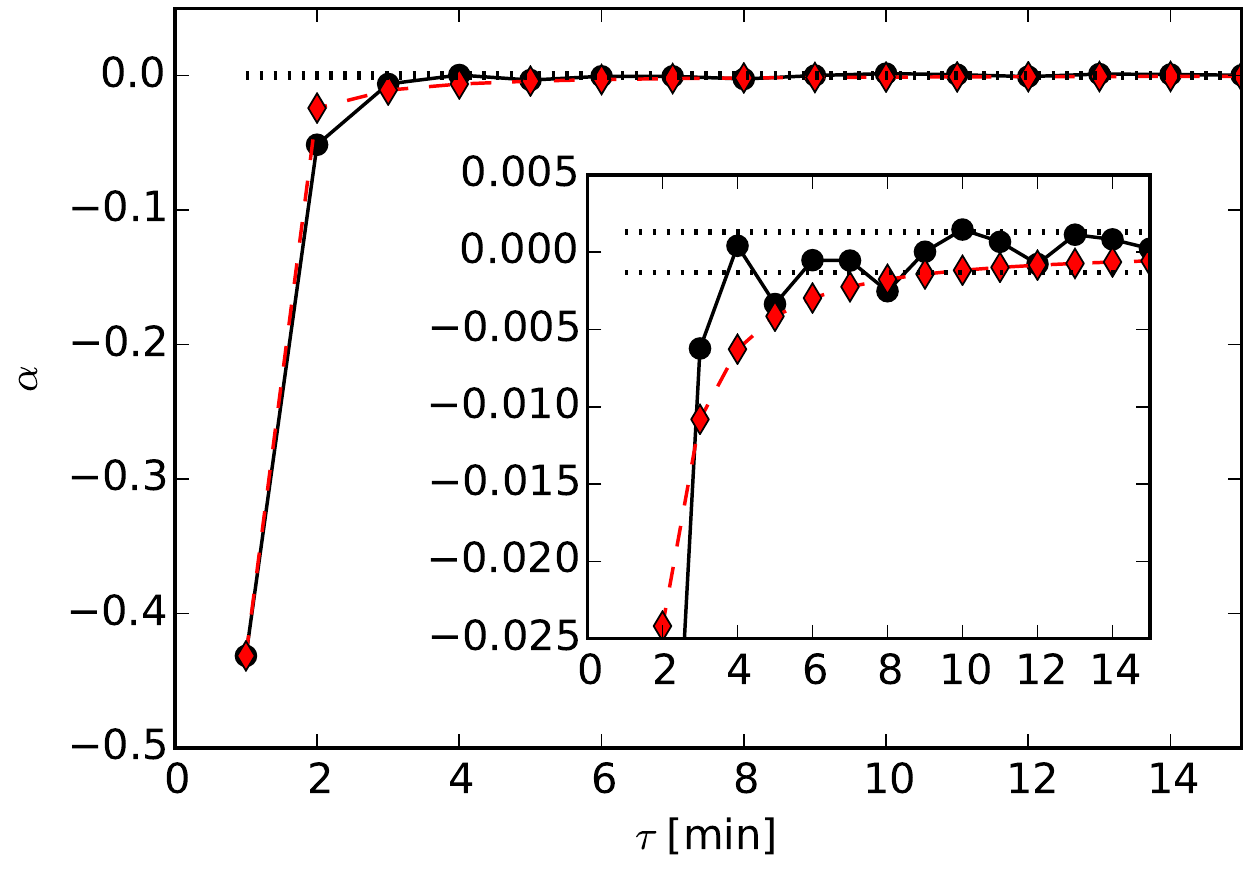}
    \caption{Autocorrelation $\alpha$ over time lag $\tau$ of one-minute increments of the flow $q$ averaged over all days and crosssections (black circles) compared with fBm with $H=0.093$ (red diamonds). Horizontal dotted lines indicate the interval spanned by shuffled data. The inset represents an enlarged part of the main figure.}
    \label{fig:auto}
\end{center}
\end{figure}

To understand the time evolution of $q(t)$, let us assume we identified the non-stochastic trend $\mu(t)$ and propose the model $\qm$, defined as \begin{multline}
    \qm(t+\Delta t)-\qm(t) = \mu(t+\Delta t)-\mu(t)\\ + \sigma(t) \left[B_H\left(\frac{t}{t_0}+\frac{\Delta t}{t_0}\right)-B_H\left(\frac{t}{t_0}\right)\right]. 
\end{multline}
The time dependent function $\sigma(t)$ is needed to adapt the physical dimension and to account for a slowly varying time dependence of the fluctuation strength.  For the time scale $t_0$ we can use one minute. 
For $H$, we insert the Hurst exponent as found empirically around $H=0.1$. For $H\neq 1/2$, the increments $B_H(\tilde{t}+\Delta \tilde{t})-B_H(\tilde{t})$ are dependent for different $\tilde{t}=n \Delta \tilde{t}$. Therefore, the numerical generation of time series is not as straight forward as for standard Brownian motion. We find a strong negative autocorrelation $\alpha(\tau)=\left[\left<\Delta q(t+\tau)\Delta q(t)\right>_t-(\left<\Delta q(t)\right>_t)^2\right]/\left<(\Delta q(t))^2\right>_t$ of one minute increments $\Delta q(t) = q(t+1)-q(t)$ for short time lags $\tau$, see the black circles in fig.~\ref{fig:auto}. Results are averaged over all road sections and all days with at most ten minutes of missing data. For larger time lags $\tau$, the autocorrelation is dominated by noise. This result is consistent with anti\-persistent fBm, as can be found with eq.~\ref{eq:dependence}. The red diamonds show results for $H=0.093$. This Hurst exponent is also in good agreement with results from DFA, see fig.~\ref{fig:dfa}. Notice that non-stochastic increments of the form $\mu(t+1)-\mu(t)$ are small compared to fluctuations on short time scales, what allows us to investigate the autocorrelation of $q$ without subtracting the non-stochastic part. 

Based on the model $\qm$, we now investigate the durations $T_{\rm thr}$ with $q_m>\qthr$. For fBm, a power law with exponent $\gamma_{\rm fBm}=2-H$ was reported in~\cite{ding1995distribution}. In our case, we find $\gamma_{\rm fBm}=2-H\approx 1.9$, which is in good agreement with the empirical findings for $\gamma_{\rm thr}$ in fig.~\ref{fig:j_duration}. For fBm, it was further found that the power law behaviour is even present with an additional drift-like term~\cite{ding1995distribution}. In this case, for negative drift there is a cut-off at large times, what is also consistent with our empirical results, see for example the red circles on the right of fig.~\ref{fig:j_duration}. For positive drift, the PDF at long durations $T$ with $\qm>\qthr$ are increased. We see this effect in real data in fig.~\ref{fig:j_duration} for small $\qthr$, the yellow circles. In our model $\qm(t)$, the drift $\mu(t)$ would be a function depending on the time of the day, the day of the week and further factors. \revision{Also, fig.~\ref{fig:j_timeseries} indicates a dependence of the fluctuation strength $\sigma(t)$ on $\mu(t)$.} However, the identification of this drift term goes beyond the scope of this study. 

Moreover, let us compare with scaling in other socio-economic fields. Burst- and inter-burst durations $T$ in currency exchange markets have been found to scale as $T^{-3/2}$ \cite{gontis2017burst}. This hints at normal diffusion and Markovian behaviour. Examples of scaling in systems which are not tuned to a phase transition are also known in the context of coherent noise \cite{newman1996avalanches}, what holds implications for adaptive electricity markets \cite{krause2015econophysics}. 

\section{Conclusion}
First, our results strongly corroborate the anti\-persistent behaviour of traffic flow time series $q(t)$: The Hurst exponent around $H=0.1$ from DFA, negative autocorrelations of one minute increments hinting at Hurst exponent around $H=0.09$, and finally a power law $T^{-\gamma_{\rm thr}}$ for durations $T_{\rm thr}$ above thresholds $q>\qthr$ with exponent around $\gamma_{\rm thr}=2$. The latter is connected with a Hurst exponent $H=2-\gamma_{\rm thr}$ close to zero, and therefore strongly in the anti\-persistent regime. Taking all findings together, we found a robust universal property of traffic flow, which can be observed on different road sections, at different times and with or without long times of congestion. 

Second, we showed that congestion durations $T$ are distributed in the same way as durations $T_{\rm thr}$ of the flow above threshold. With identifying critical thresholds of the flow $\qthr$ for our traffic data, we concluded that the durations $T_{\rm thr}$ translate into traffic jam durations $T$. 

This led us, third, to our main result that anti\-persistence in traffic flow is a crucial property for understanding patterns of traffic congestion. \revision{The fact that the traffic flow can be described with a fractional Brownian motion, with a subtle time dependence of fluctuations, and that it strongly influences patterns of traffic breakdown, implies a broad distribution of congestion lifetimes. Especially for antipersistent fractional Brownian motion, the role of short lasting jams is increased. Accordingly we found that short jams of duration $T\leq 10 \un{min}$ contribute 19\% to the total sum of jam hours.} This is relevant for navigation systems with congestion warning. Especially short lasting traffic jams bare a large risk for rear-end collisions. Also, traffic models can benefit from our findings. 

\acknowledgments
We thank Strassen.NRW for providing the empirical traffic data. 
LH and MS have been supported by Deutsche Forschungsgemeinschaft (DFG) within the Collaborative Research Center SFB 876 ``Providing Information by Resource-Constrained Analysis'', project B4 ``Analysis and Communication for the Dynamic Traffic Prog\-nosis''.
\bibliography{ptt-references}

\end{document}